\documentclass[twocolumn,pre,superscriptaddress,showpacs]{revtex4-1}

\usepackage{url}
\usepackage{graphicx}
\usepackage{epstopdf}
\usepackage{algorithm}
\usepackage{algorithmicx}
\usepackage{algpseudocode}
\usepackage{calrsfs}
\usepackage{color}
\usepackage{amsmath,amssymb,amsfonts,amsthm}
\usepackage{bm}
\usepackage{setspace}
\usepackage{multirow}
\usepackage{mathtools}

\newcommand{\asconv}{\overset{\text{a.s.}}{\longrightarrow}}
\newcommand{\ra}{\rightarrow}
\newcommand{\LB}{\left\{}
\newcommand{\RB}{\right\}}
\newcommand{\Lb}{\left[}
\newcommand{\Rb}{\right]}
\newcommand{\lb}{\left(}
\newcommand{\rb}{\right)}

\newcommand{\pone}{p_1}
\newcommand{\ptwo}{p_2}
\newcommand{\none}{{n_1}}
\newcommand{\ntwo}{{n_2}}
\newcommand{\hnone}{\widehat{n}_1}
\newcommand{\hntwo}{\widehat{n}_2}
\newcommand{\Aone}{\mathbf{A}_1}
\newcommand{\Atwo}{\mathbf{A}_2}
\newcommand{\bDelta}{\mathbf{\Delta}}
\newcommand{\bA}{\mathbf{A}}

\newcommand{\bC}{\mathbf{C}}
\newcommand{\bCbar}{\mathbf{\bar{C}}}

\newcommand{\bone}{\mathbf{1}}
\newcommand{\onenone}{\mathbf{1}_{\none}}
\newcommand{\onentwo}{\mathbf{1}_{\ntwo}}

\newcommand{\yone}{\mathbf{y}_1}
\newcommand{\ytwo}{\mathbf{y}_2}
\newcommand{\bd}{\mathbf{d}}
\newcommand{\done}{\mathbf{d}_1}
\newcommand{\dtwo}{\mathbf{d}_2}
\newcommand{\bdt}{\mathbf{\widetilde{d}}}
\newcommand{\dtone}{\mathbf{\widetilde{d}}_1}
\newcommand{\dttwo}{\mathbf{\widetilde{d}}_2}
\newcommand{\xone}{\mathbf{x}_1}
\newcommand{\xtwo}{\mathbf{x}_2}
\newcommand{\bx}{\mathbf{x}}
\newcommand{\by}{\mathbf{y}}
\newcommand{\bz}{\mathbf{z}}

\newcommand{\hp}{\widehat{p}}
\newcommand{\hpstar}{\widehat{p^*}}
\newcommand{\hpone}{\widehat{p}_1}
\newcommand{\hptwo}{\widehat{p}_2}

\newcommand{\bB}{\mathbf{B}}
\newcommand{\Bone}{\mathbf{B}_1}
\newcommand{\Btwo}{\mathbf{B}_2}

\begin{document}
\title{Universal Phase Transition in Community Detectability under a Stochastic Block Model}

\author{
Pin-Yu Chen and Alfred O. Hero III\\
Department of Electrical Engineering and Computer Science\\
University of Michigan,
Ann Arbor, MI 48109, USA \\
}

\begin{abstract}
We prove the existence of an asymptotic phase transition threshold on community detectability for the spectral modularity method
[M. E. J. Newman, Phys. Rev. E 74, 036104 (2006) and Proc. National Academy of Sciences. 103, 8577 (2006)] under a stochastic block model.
The phase transition on community detectability occurs as the inter-community edge connection probability $p$ grows.
This phase transition separates a sub-critical regime of small $p$, where modularity-based community detection successfully identifies the communities, from a super-critical regime of large $p$ where successful community detection is impossible. We show that, as the community sizes become large, the asymptotic phase transition threshold $p^*$ is equal to $\sqrt{\pone \ptwo}$, where $p_i~(i=1,2)$ is the within-community edge connection probability. Thus the phase transition threshold is universal in the sense that it does not depend on the ratio of community sizes. The universal phase transition phenomenon is validated by simulations for moderately sized communities. Using the derived expression for the phase transition threshold we propose an empirical method for estimating this threshold from real-world data.
\end{abstract}

\pagestyle{empty}
\maketitle
\small

\section{Introduction}
\label{sec_Intro}
Community detection is an active research field that arises in technological, social, and biological networks. The goal of community detection is to detect tightly
connected subgraphs in a graph \cite{Fortunato10}. The spectral modularity method proposed by Newman \cite{Newman06community,Newman06PNAS} is widely applied to community detection.
It has been observed that community detectability (i.e., the fraction of correctly identified nodes) degrades rapidly as the number of inter-community edges increases beyond a certain critical value \cite{Bickel09,Ronhovde09,Decelle11,Decelle11PRE,Zhang12,Nadakuditi12Detecability,Krzakala13,Radicchi13_hetero,Radicchi14}.
This paper establishes a mathematical expression for the critical phase transition threshold in modularity-based community detection under a stochastic block model. This phase transition threshold governs the community modularity measure of the graph as a function of the respective edge connection probabilities $p_1$ and $p_2$ within community $1$ and community $2$. Defining $p$ as the edge connection probability between the two communities the critical phase transition threshold on $p$ takes on the simple asymptotic form
$p^*=\sqrt{p_1p_2}$, in the limit as the two community sizes converge (at comparable rate) to infinity. Remarkably, $p^*$ does not depend on the community sizes, and in this sense it is a universal threshold.

Let $n$ denote the total number of nodes in an undirected graph and let $\bA$ be the associated adjacency matrix. Specifically, $\bA$ is
an $n \times n$ binary symmetric matrix characterizing the connectivity structure of a graph, where
$\bA_{ij}=1$ if an edge exists between node $i$ and node $j$, and $\bA_{ij}=0$ otherwise.
Newman proposed a measure called modularity that evaluates the number of excessive edges of a graph compared with the corresponding degree-equivalent random graph. More specifically, define the modularity matrix as $\mathbf{B}=\bA-b \mathbf{d}\mathbf{d}^T$, where $\mathbf{d}$ is the degree vector of the graph and $b$ is the reciprocal of the total number of edges in the graph.
The last term $b \mathbf{d}\mathbf{d}^T$ can be viewed as the expected adjacency matrix of the degree-equivalent random graph.
Newman proposed to compute the largest eigenvector of $\bB$ and perform K-means clustering \cite{Hartigan1979} or take the sign function on this vector to cluster the nodes
into two communities. Since the $n$-dimensional vector of all ones, $\bone_n$, is always in the null space of $\bB$, i.e, $\bB \bone_n=\mathbf{0}_n$, where $\mathbf{0}_n$ is the $n$-dimensional vector of all zeros, the (unnormalized) modularity is the largest eigenvalue of $\bB$ and has the representation
\begin{align}
\label{eqn_modularity_def}                                                                                                                                                   \lambda_{\max}(\bB)= \max_{\bx^T \bx=1,~\bx^T \bone_n=0} \bx^T \bB \bx.
\end{align}

Consider a stochastic block model \cite{Holland83} consisting of two community structures
parameterized by edge connection probability $p_i$ within community $i$  ($i=1,2$) and edge connection probability $p$
between the two communities. Let $n_i$ denote the size of community $i$
such that $\none+\ntwo=n$. The overall $n \times n$ adjacency matrix of the entire graph can be
represented as
\begin{align}                                                                                                                                                                      \label{eqn_asym_block_model}
\mathbf{A} = \begin{bmatrix}                                                                                                                                                       \Aone & \bC           \\
\bC^T       & \Atwo
\end{bmatrix},                                                                                                                                                                \end{align}
where $\mathbf{A}_i$ is the $n_i$-by-$n_i$ adjacency matrix of an Erdos-Renyi random graph with edge connection probability $p_i$ and $\bC$ is the $n_1$-by-$n_2$ adjacency
matrix of the inter-community edges where each entry in $\bC$ is a Bernoulli($p$) random variable.
A similar network model is studied in \cite{Radicchi13} for interconnected networks.
However, in \cite{Radicchi13} the communities (subnetworks) have the same size and the inter-community edges are known (i.e., non-random).
The main purpose of \cite{Radicchi13} is to study the eigenstructure of the overall graph Laplacian matrix with different interconnected edge strengths, as contrasted to community detection. In \cite{CPY14spectral}, the network model (\ref{eqn_asym_block_model}) is used to study community detectability of spectral algorithms based on the eigenvectors of the graph Laplacian matrix.

The fundamental limits on community detectability have been investigated for the spectral modularity method under
more restrictive assumptions \cite{Nadakuditi12Detecability,Decelle11} than assumed in this paper.
In \cite{Nadakuditi12Detecability}, the community detectability of the spectral modularity method is studied in sparse random networks
where the average degree is fixed and the two communities have the same community size and identical within-community edge connection probability, i.e., $\none=\ntwo$,
$\pone=\ptwo=O(\frac{1}{n})$, and $p=O(\frac{1}{n})$. The critical value for community detectability is shown to depend on the average degree of the within-community and inter-community edges.
Similar closed-form phase transition expressions have been found under the same network assumption in \cite{Decelle11,Krzakala13,Decelle11PRE}.

The planted clique detection problem in \cite{Nadakuditi12plant} is a further restriction of the stochastic block model when $p_2=p$.
For spectral methods that use the eigenvectors of linear operators associated with the graph for community detection (e.g., the modularity, adjacency, Laplacian, or normalized Laplacian matrices), the phase transition threshold under the general stochastic block model can be derived by investigating the eigenvalue spectra \cite{Peixoto13}.

Different from the aforementioned works, our network model relaxes the assumptions of identical community size and within-community edge connection probability,
and we assume that the parameters $\pone$ and $\ptwo$ are fixed. Under this general setting, we prove an asymptotic universal phase transition threshold
of $p$ on community detection using the spectral modularity method, where the asymptotic critical value of $p$ is $p^*=\sqrt{\pone \ptwo}$.
We also derive asymptotic forms for the modularity and the largest eigenvector of $\bB$, which are directly affected by the phase transition phenomenon. Note that the same phase transition threshold has been derived in \cite{Zhao12} in terms of the consistency of the modularity and the loglikelihood of the degree corrected stochastic model \cite{Karrer11}, whereas in this paper we explicitly show that the spectral modularity method can achieve the same phase transition threshold. Also note that under the same stochastic block model (\ref{eqn_asym_block_model}), the phase transition threshold of the spectral modularity method established in this paper coincides with
the phase transition threshold of several spectral community detection methods derived from Eq. (6) in \cite{Peixoto13}. This suggests that this phase transition threshold might be universal for many spectral methods.

\section{Phase Transition Analysis}
Using the network model in (\ref{eqn_asym_block_model}), let $\bd=\bA \bone_n=[\done^T~\dtwo^T]^T$ denote the degree vector of the graph with $\done \in \mathbb{R}^\none$
and $\dtwo \in \mathbb{R}^\ntwo$. Then $b=(\bone_n^T \bA \bone_n)^{-1}=(\done^T \onenone + \dtwo^T \onentwo)^{-1}$.
Let $\bdt_i=\bA_i \bone_{n_i}$ denote the degree vector of community $i$. Since $\bA \bone_n = \bd$,
with $(\ref{eqn_asym_block_model})$ the degree vectors $\done$, $\dtwo$, $\dtone$, and $\dttwo$ satisfy the following equations:
\begin{align}
\label{eqn_degree_relation}
\done = \dtone + \bC \onentwo
~~\text{and}~~
\dtwo = \dttwo + \bC^T \onenone.
\end{align}
Let $b_i=(\bdt_i^T \bone_{n_i})^{-1}$. The modularity matrix of community $i$ is denoted by $\bB_i=\bA_i-b_i \bdt_i \bdt_i^T$. Using these notations, the modularity matrix of the entire graph can be represented as
\begin{align}
\label{eqn_modularity_block}
\bB = \begin{bmatrix}
       \Bone+b_1 \dtone \dtone^T -b \done \done^T & \bC-b \done \dtwo^T           \\
       \bC^T-b \dtwo \done^T           & \Btwo+b_2 \dttwo \dttwo^T-b \dtwo \dtwo^T
     \end{bmatrix}.
\end{align}

Let $\by=[\yone^T~\ytwo^T]^T$ denote the largest eigenvector of $\bB$, where $\yone \in \mathbb{R}^\none$ and $\ytwo \in \mathbb{R}^\ntwo$.
Following the definition of modularity in (\ref{eqn_modularity_def}) and (\ref{eqn_modularity_block}),
$\by =\arg \max_{\bx} \Gamma(\bx)$, where
\begin{align}
\label{eqn_Lagrange_modularity}
\Gamma(\bx)&=\xone^T \Bone \xone + \xtwo^T \Btwo \xtwo + b_1(\dtone^T \xone)^2  + b_2 (\dttwo^T \xtwo)^2 \nonumber \\
&~~~-b(\done^T \xone)^2-b(\dtwo^T \xtwo)^2 +2 \xone^T \bC \xtwo
- 2b (\done^T \xone) (\dtwo^T \xtwo) \nonumber \\
&~~~- \mu (\xone^T \xone + \xtwo^T\xtwo -1)-\nu (\xone^T\onenone+\xtwo^T \onentwo),
\end{align}
and $\bx=[\xone^T~\xtwo^T]^T$, $\xone \in \mathbb{R}^\none$, and $\xtwo \in \mathbb{R}^\ntwo$. $\mu$ and $\nu$ are Lagrange multipliers of the constraints $\bx^T \bx=1$ and $\bx^T \bone_n=0$ in (\ref{eqn_modularity_def}), respectively.

Differentiating (\ref{eqn_Lagrange_modularity}) with respect to $\xone$ and $\xtwo$ respectively, and substituting $\by$ to the equations, we obtain
\begin{align}
\label{eqn_modularity_DE1}
&2 \Bone \yone + 2b_1 (\dtone^T \yone) \dtone - 2b (\done^T \yone) \done  - 2b(\dtwo^T \ytwo) \done \nonumber \\
&~~~+ 2 \bC \ytwo - 2 \mu \yone -\nu \onenone=\mathbf{0}_{\none}; \\
\label{eqn_modularity_DE2}
&2 \Btwo \ytwo + 2b_2 (\dttwo^T \ytwo) \dttwo - 2b (\dtwo^T \ytwo) \dtwo  - 2b(\done^T \yone) \dtwo \nonumber \\
&~~~+ 2 \bC^T \yone - 2 \mu \ytwo -\nu \onentwo=\mathbf{0}_{\ntwo}.
\end{align}
Left multiplying (\ref{eqn_modularity_DE1}) by $\onenone^T$  and left multiplying (\ref{eqn_modularity_DE2}) by $\onentwo^T$  and
recalling that $\bB_i \bone_{n_i} = \mathbf{0}_{n_i}$ and $b_i=(\bdt_i^T \bone_{n_i})^{-1}$, we have
\begin{align}
\label{eqn_modularity_DE1_by_one}
&2 (\dtone^T \yone)-2b (\done^T \yone)(\done^T \onenone) - 2b (\dtwo^T \ytwo) (\done^T \onenone) + 2 \onenone^T \bC \ytwo   \nonumber \\
&~~~-2 \mu \yone^T \onenone - \nu \none=0; \\
\label{eqn_modularity_DE2_by_one}
&2 (\dttwo^T \ytwo)-2b (\dtwo^T \ytwo)(\dtwo^T \onentwo) - 2b (\done^T \yone) (\dtwo^T \onentwo) + 2 \onentwo^T \bC^T \yone \nonumber \\
&~~~-2 \mu \ytwo^T \onentwo - \nu \ntwo=0.
\end{align}
Summing (\ref{eqn_modularity_DE1_by_one}) and (\ref{eqn_modularity_DE2_by_one}) and using (\ref{eqn_degree_relation}) gives $\nu=0$.
Left multiplying (\ref{eqn_modularity_DE1}) by $\yone^T$ and left multiplying (\ref{eqn_modularity_DE2}) by $\ytwo^T$, substituting $\nu=0$ and summing
the equations, with
(\ref{eqn_modularity_block}) we have $\mu=\lambda_{\max}(\bB)$.

Let $\bCbar=p \onenone \onentwo^T$, a matrix whose elements are the means of entries in $\bC$. Let $\sigma_i(\mathbf{M})$ denote the $i$-th largest singular value of a rectangular matrix $\mathbf{M}$ and write $\bC=\bCbar+\bDelta$, where $\bDelta=\bC-\bCbar$. Latala's theorem \cite{Latala05} implies that the expected value of $\sigma_1 \lb \frac{\bDelta}{\sqrt{\none \ntwo}}\rb$ converges to $0$ as $\none$ and $\ntwo$ approach to infinity, denoted $\mathbb{E} \Lb \sigma_1\lb \frac{\bDelta}{\sqrt{\none \ntwo}} \rb \Rb \ra 0$ as $\none,\ntwo \ra \infty$.
This is proved in Appendix \ref{appen_Latala}.
Furthermore, by Talagrand's concentration theorem \cite{Talagrand95},
\begin{align}
\label{eqn_Talagrand}
 \sigma_1\lb \frac{\bC}{\sqrt{\none \ntwo}} \rb \asconv  p
\text{~~and~~}
 \sigma_i \lb \frac{\bC}{\sqrt{\none \ntwo}} \rb \asconv 0,~~\forall i \geq 2
\end{align}
when $\none,\ntwo \ra \infty$, where $\asconv$ means almost sure convergence. This is proved in Appendix \ref{appen_Talagrand}. Note that the convergence rate is maximal when $\none=\ntwo$ because $\none+\ntwo \geq 2\sqrt{\none \ntwo}$ and the equality holds if $\none=\ntwo$.

Throughout this paper we further assume $\frac{\none}{\ntwo} \ra c >0$ as $\none,\ntwo \ra \infty$. This means the community sizes grow with comparable rates.
As proved in \cite{BenaychGeorges12}, the singular vectors of $\bC$ and $\bCbar$ are close to each other in the sense that the square of inner product of their left/right singular vectors converges to $1$ almost surely when $\sqrt{\none \ntwo} p \rightarrow \infty$. Consequently, the concentration results in (\ref{eqn_Talagrand}) and \cite{BenaychGeorges12} imply that
\begin{align}
\label{eqn_C_concentration}
\frac{\bC \onentwo}{\ntwo} \asconv p \onenone
\text{~~and~~}
\frac{\bC^T \onenone}{\none} \asconv p \onentwo.
\end{align}

Furthermore, since under the stochastic block model setting
each entry of the adjacency matrix $\bA_i$ in (\ref{eqn_asym_block_model}) is a Bernoulli($p_i$) random variable,
following the same concentration arguments in (\ref{eqn_Talagrand}) and (\ref{eqn_C_concentration}) we have
\begin{align}
\label{eqn_A_concentration_SBM}
\frac{\Aone \onenone}{\none} \asconv \pone \onenone
\text{~~and~~}
\frac{\Atwo \onentwo}{\ntwo} \asconv \ptwo \onentwo.
\end{align}
By the fact that $\bdt_i=\bA_i \bone_{n_i}$, (\ref{eqn_A_concentration_SBM}) implies that
\begin{align}
\label{eqn_dtilde_concentration}
\frac{\dtone}{\none} \asconv \pone \onenone~~\text{and}~~\frac{\dttwo}{\ntwo} \asconv \ptwo \onentwo.
\end{align}
Applying (\ref{eqn_C_concentration}), (\ref{eqn_A_concentration_SBM}) and (\ref{eqn_dtilde_concentration}) to (\ref{eqn_degree_relation})
and recalling that $\frac{\none}{\ntwo} \ra c>0$, we have
\begin{align}
\label{eqn_degree_relation_concentration}
\frac{\done}{\none} \asconv \lb \pone + \frac{p}{c}  \rb \onenone
~~\text{and}~~
\frac{\dtwo}{\ntwo} \asconv \lb \ptwo   + c p \rb \onentwo.
\end{align}
Therefore the reciprocal of the total degree in the graph $b$ has the relation
\begin{align}
\label{eqn_b_concentration}
\none \ntwo b=\frac{\none \ntwo}{\done^T \onenone+\dtwo^T \onentwo} \asconv \frac{1}{c \pone+2p+\frac{\ptwo}{c}}.
\end{align}
Substituting these limits to (\ref{eqn_modularity_DE1_by_one}) and (\ref{eqn_modularity_DE2_by_one}) and recalling that $\nu=0$ and $\yone^T \onenone=-\ytwo^T \onentwo$,
we have
\begin{align}
\label{eqn_modularity_SBM_condition_1}
&\yone^T\onenone \lb \frac{\mu}{n}-\frac{\pone \ptwo -p^2}{c \pone+2p+\frac{\ptwo}{c}}\rb \asconv 0;   \\
\label{eqn_modularity_SBM_condition_2}
&\ytwo^T\onentwo \lb \frac{\mu}{n}-\frac{\pone \ptwo -p^2}{c \pone+2p+\frac{\ptwo}{c}}\rb \asconv 0.
\end{align}
Since $\mu=\lambda_{\max}(\bB)$, for each inter-community edge connection probability $p$, one of the two cases below has to be satisfied:
\begin{align}
\label{eqn_tranisition_condition_1}
&\text{Sub-critical regime:~}\frac{\lambda_{\max}(\bB)}{n} \asconv \frac{\pone \ptwo -p^2}{c \pone+2p+\frac{\ptwo}{c}}   \\
\label{eqn_tranisition_condition_2}
&\text{Super-critical regime:~}\yone^T\onenone \asconv 0~~\text{and}~~\ytwo^T\onentwo \asconv 0
\end{align}

In the sub-critical regime, observe that $\frac{\lambda_{\max}(\bB)}{n}$ converges to $\frac{\pone \ptwo -p^2}{c \pone+2p+\frac{\ptwo}{c}}$ almost surely such that the corresponding
asymptotic largest eigenvector $\by$ of $\bB$ remains the same (unique up to its sign) for different $p$.
Left multiplying  (\ref{eqn_modularity_DE1}) by $\yone^T$ and  left multiplying (\ref{eqn_modularity_DE2}) by $\ytwo^T$, summing these two equations,
and using the limiting expressions (\ref{eqn_modularity_block}), (\ref{eqn_C_concentration}), (\ref{eqn_A_concentration_SBM}), (\ref{eqn_dtilde_concentration}), (\ref{eqn_degree_relation_concentration}), (\ref{eqn_b_concentration}), and (\ref{eqn_tranisition_condition_1}), in the sub-critical regime, we have
\begin{align}
\label{eqn_modularity_eigvector_concentration}
\frac{\yone^T \Bone \yone }{n}+  \frac{\ytwo^T \Btwo \ytwo }{n}+ f(p) \asconv 0,
\end{align}
where $f(p)=\frac{\pone \ptwo - p^2}{c \pone+2p+\frac{\ptwo}{c}} \Lb \frac{\lb \sqrt{c}+\frac{1}{\sqrt{c}}\rb^2 \lb \yone^T \onenone \rb^2}{n} -1 \Rb$.
Since $f(p)$ is a Laurent polynomial of $p$ with finite powers, and (\ref{eqn_modularity_eigvector_concentration}) has to be satisfied over all values of $p$ in the sub-critical regime,
\begin{align}
\label{eqn_modularity_eigvector_concentration_2}
\frac{\yone^T \Bone \yone }{n}+  \frac{\ytwo^T \Btwo \ytwo }{n} \asconv 0~~\text{and}~~f(p) \asconv 0.
\end{align}
Furthermore, we can show that, in the sub-critical regime, $\yone$ and $\ytwo$ converge almost surely to constant vectors with opposite signs,
\begin{align}
\label{eqn_modularity_eigvector_concentration_3}
\sqrt{\frac{n \none}{\ntwo}} \yone \asconv \pm \onenone
~~\text{and}~~
\sqrt{\frac{n \ntwo}{\none}} \ytwo \asconv \mp \onentwo.
\end{align}
This is proved in Appendix \ref{appen_eigenvector_convergence}.
Therefore, in the sub-critical regime the two communities can be almost perfectly detected. On the other hand, in the super-critical regime the spectral modularity method fails to detect the two communities
since by (\ref{eqn_tranisition_condition_2}) $\yone$ and $\ytwo$ must have both positive and negative entries.

Next we derive the asymptotic universal phase transition threshold $p^*$ for transition from the sub-critical regime to the super-critical regime that occurs as $p$ increases. Note that in the super-critical regime, since
$\yone^T\onenone \asconv 0$ and $\ytwo^T\onentwo \asconv 0$, using (\ref{eqn_modularity_def}), (\ref{eqn_modularity_block}), (\ref{eqn_C_concentration}), (\ref{eqn_A_concentration_SBM}), (\ref{eqn_dtilde_concentration}) and (\ref{eqn_degree_relation_concentration})
we have
\begin{widetext}
\begin{align}
\label{eqn_phase_transition_case_2}
\frac{\lambda_{\max}(\bB)}{n} &=\frac{1}{n} \Lb \yone^T \Bone \yone + \ytwo^T \Btwo \ytwo + b_1(\dtone^T \yone)^2  + b_2 (\dttwo^T \ytwo)^2-b(\done^T \yone)^2-b(\dtwo^T \ytwo)^2 +2 \yone^T \bC \ytwo - 2b (\done^T \yone) (\dtwo^T \ytwo) \Rb \nonumber \\
&\asconv \frac{1}{n} \LB \yone^T ( \pone \onenone \onenone ^T - \pone \onenone \onenone ^T ) \yone
+ \ytwo^T ( \ptwo \onentwo \onentwo ^T - \ptwo \onentwo \onentwo^T ) \ytwo + b_1( \none \pone  \yone^T \onenone)^2  + b_2 (\ntwo \ptwo \ytwo^T \onentwo)^2 \right. \nonumber \\
&~~~~~~~~~-b \Lb \lb \none \pone+\ntwo p \rb \yone^T \onenone \Rb^2-b \Lb \lb \ntwo \ptwo+\none p \rb \ytwo^T \onentwo \Rb^2
+2 p (\yone^T \onenone)  (\ytwo^T \onentwo) \nonumber \\
&~~~~~~~~~\left.-2b \Lb \lb \none \pone+\ntwo p \rb \yone^T \onenone \Rb \Lb \lb \ntwo \ptwo+\none p \rb \ytwo^T \onentwo \Rb \RB \nonumber \\
&=0.
\end{align}                                                                                                                                                                       \end{widetext}
Consequently, by (\ref{eqn_tranisition_condition_1}) and (\ref{eqn_phase_transition_case_2}), the phase transition occurs at $p=p^*$ almost surely when
$\frac{\pone \ptwo -{p^*}^2}{c \pone+2p^*+\frac{\ptwo}{c}}=0$. This implies an asymptotic universal phase transition threshold on community detectability:
\begin{align}
\label{eqn_modularity_universal_phase_transition_SBM}
p^* \asconv \sqrt {\pone \ptwo}
\end{align}
as $\none,\ntwo \ra \infty$ and $\frac{\none}{\ntwo} \ra c >0$.
Note that the limit (\ref{eqn_modularity_universal_phase_transition_SBM}) does not depend on the community sizes. In this sense, the phase transitions are universal as they only depend on the within-community connection probabilities $p_1$ and $p_2$.

Moreover, the same phase transition results hold for a more general setting where $p_i=O(\frac{1}{n^{\epsilon}})$ and $p=O(\frac{1}{n^{\epsilon}})$ for any $\epsilon \in [0,1)$ by following the same derivation procedures. As a comparison, the phase transition threshold under the sparse network setting, where $p_i=O(\frac{1}{n})$ and $p=O(\frac{1}{n})$ \cite{Decelle11,Decelle11PRE,Zhang12,Nadakuditi12Detecability,Radicchi13_hetero,Krzakala13}, is different from the threshold established in this paper where $p_i=O(\frac{1}{n^{\epsilon}})$ and $p=O(\frac{1}{n^{\epsilon}})$ for any $\epsilon \in [0,1)$. Also note that when $p_i=O(\frac{1}{n^{\epsilon}})$ and $p=O(\frac{1}{n^{\epsilon}})$ for any $\epsilon \in [0,1)$, the community detectability undergoes an abrupt transition at the threshold whereas the transition is more smooth for sparse networks.

\begin{figure}[!t]
    \centering
    \includegraphics[width=3.7in]{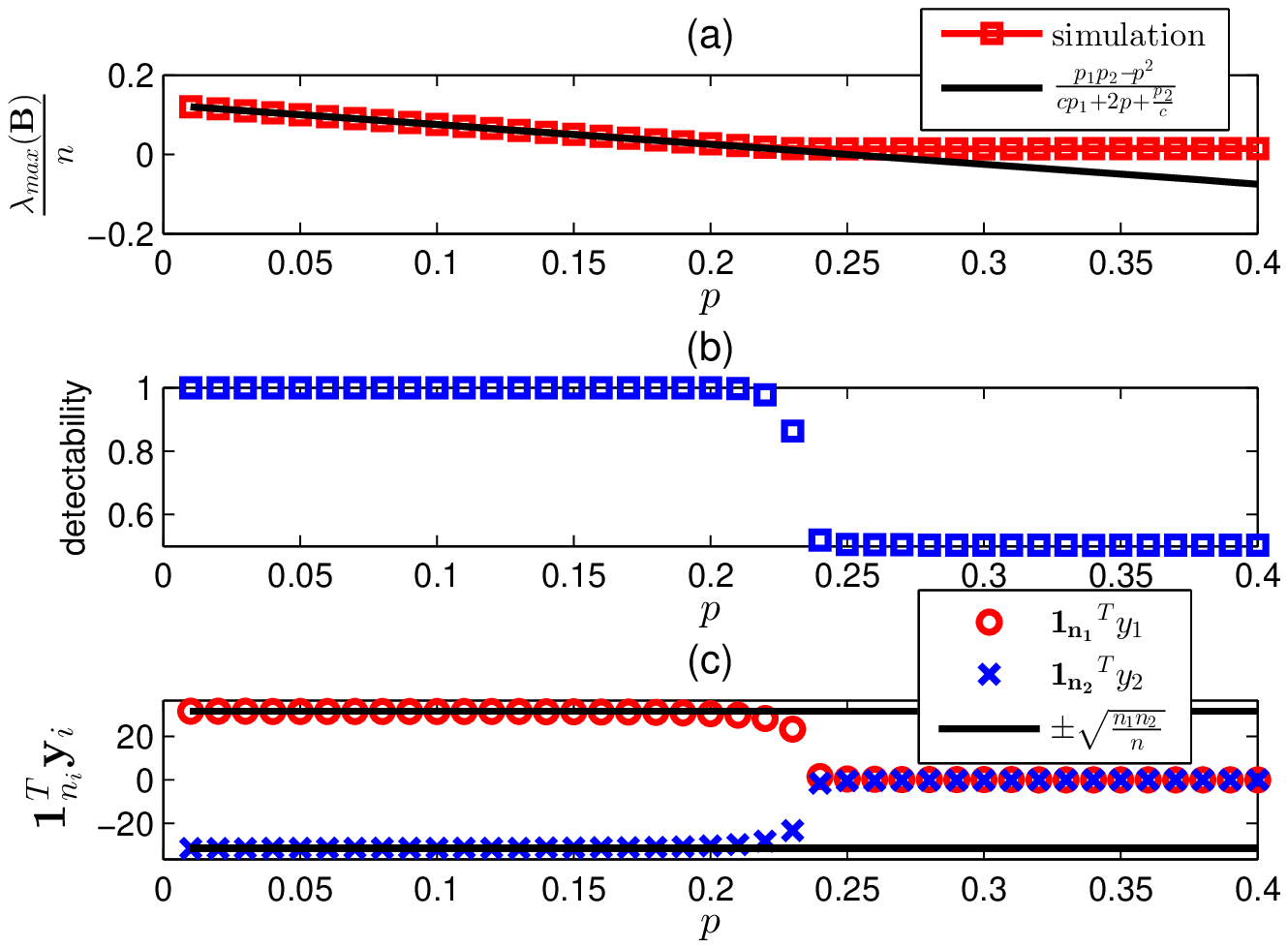}
    \caption{Validation of theoretical critical phase transition threshold (\ref{eqn_modularity_universal_phase_transition_SBM}) for two communities generated by a stochastic block model. The curves represent averages over $100$ realizations of the model. Here $\none=\ntwo=2000$ and $\pone=\ptwo=0.25$ so that the predicted critical phase transition is  $p^*=0.25$.
(a) When $p < p^*$, $\frac{\lambda_{\max}(\bB)}{n}$ converges to $\frac{\pone \ptwo -p^2}{c \pone+2p+\frac{\ptwo}{c}}$
as predicted in (\ref{eqn_tranisition_condition_1}). When $p > p^*$,  $\frac{\lambda_{\max}(\bB)}{n}$ converges to $0$ as predicted by (\ref{eqn_phase_transition_case_2}).
(b) Fraction of nodes that are correctly identified using the spectral modularity method. Community detectability undergoes a phase transition from perfect detectability to low detectability at $p=p^*$.
(c) The spectral modularity method fails to detect the communities when $p > p^*$ since the components of the largest eigenvector of $\bB$, $\yone$ and $\ytwo$,
undergo transitions at $p =p^*$ as predicted by (\ref{eqn_tranisition_condition_2}) and (\ref{eqn_modularity_eigvector_concentration_3}).
}
    \label{Fig_modularity_2000_2000_500_500}
\end{figure}

\begin{figure}[!t]
    \centering
    \includegraphics[width=3.7in]{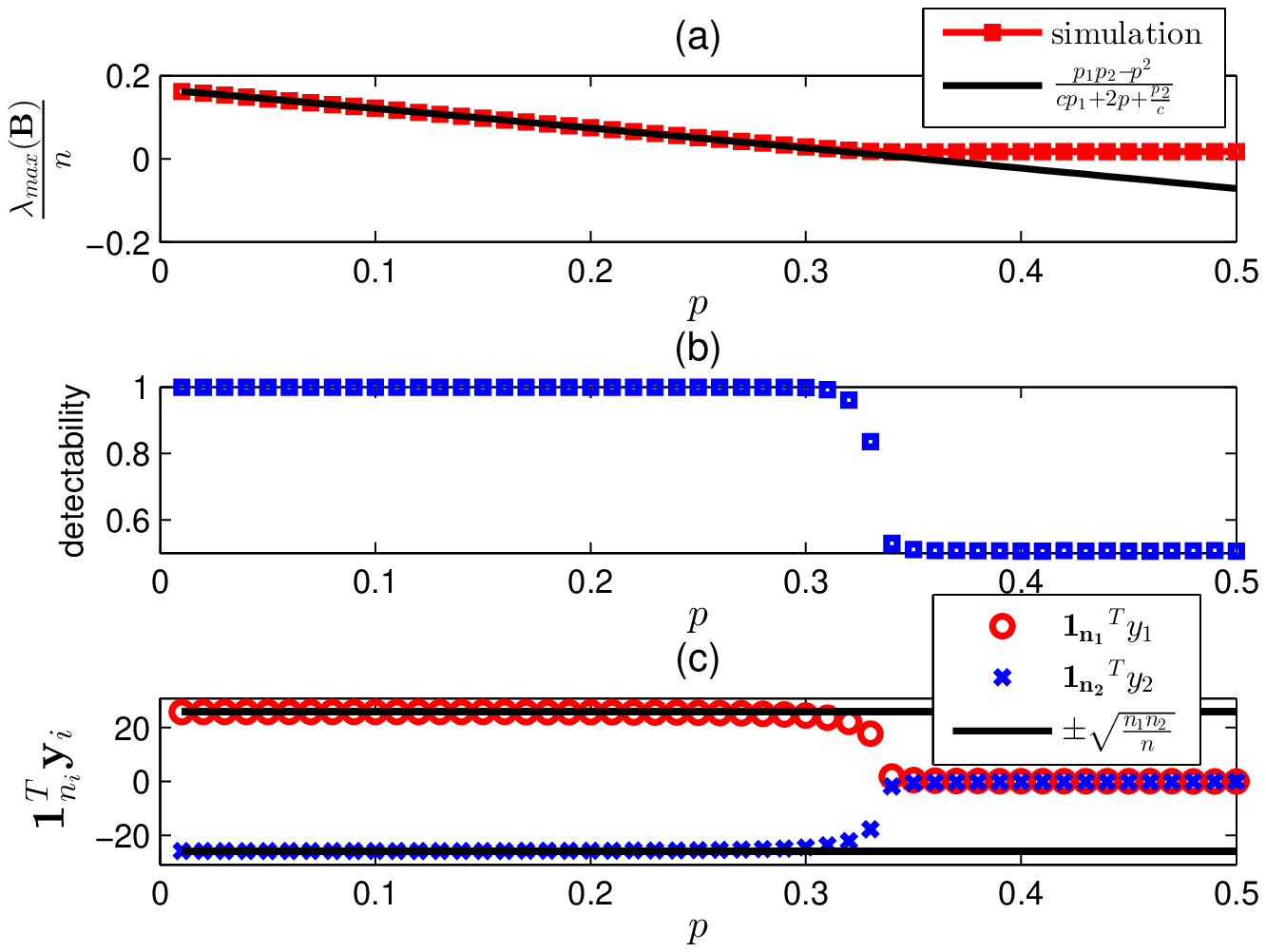}
\caption{Validation of theoretical critical phase transition threshold (\ref{eqn_modularity_universal_phase_transition_SBM}) for two communities generated by a stochastic block model. The curves represent averages over $100$ realizations of the model. Here $\none=1000$, $\ntwo=2000$, $\pone=0.5$, and $\ptwo=0.25$  so that the predicted critical phase transition is $p^*=0.3536$.
Similar phase transition phenomenon can be observed for this network setting.}
\label{Fig_modularity_1000_2000_500_500}
\end{figure}

\section{Performance Evaluation}

\subsection{Numerical Results}

We validate the asymptotic phase transition phenomenon predicted by our theory, and in particular the critical phase transition threshold (\ref{eqn_modularity_universal_phase_transition_SBM}), showing that the asymptotic theory provides remarkably accurate predictions for the case of finite small community sizes.
Fig. \ref{Fig_modularity_2000_2000_500_500} (a) shows that $\frac{\lambda_{\max}(\bB)}{n}$ converges to $\frac{\pone \ptwo -p^2}{c \pone+2p+\frac{\ptwo}{c}}$ when $p < p^*$ and
$\frac{\lambda_{\max}(\bB)}{n}$ converges to $0$ when  $p > p^*$, as predicted by (\ref{eqn_modularity_SBM_condition_1}) and (\ref{eqn_phase_transition_case_2}).
Fig. \ref{Fig_modularity_2000_2000_500_500} (b) shows the phase transition from perfect
detectability to low detectability at the critical value $p=p^*$. The numerical phase transition thresholds are
accurately predicted by (\ref{eqn_modularity_universal_phase_transition_SBM}).
Fig. \ref{Fig_modularity_2000_2000_500_500} (c) further validates the predictions in (\ref{eqn_tranisition_condition_2})
 and (\ref{eqn_modularity_eigvector_concentration_3}) that $\yone$ and $\ytwo$ converge almost surely to constant vectors with opposite signs in the sub-critical regime of $p<p^*$ and $\yone^T \onenone$ and $\ytwo^T \onentwo$ converge to $0$ almost surely in the super-critical regime of $p > p^*$. Similarly in Fig. \ref{Fig_modularity_1000_2000_500_500}, the results are shown for a different stochastic block model
 where the sizes of the two communities are not the same. These results validate that the asymptotic phase transition threshold $p^*$ in (\ref{eqn_modularity_universal_phase_transition_SBM}) is a universal phenomenon that does not depend on the community sizes.
 We have observed (see Appendix \ref{appen_finite_size}) that the asymptotic phase transition expression in (\ref{eqn_modularity_universal_phase_transition_SBM}) is accurate even in cases of relatively small community sizes, e.g. down to sizes as small as $100$.

\subsection{Empirical Estimator of the Phase Transition Threshold}
Using the derived expression of the phase transition threshold in (\ref{eqn_modularity_universal_phase_transition_SBM}), we propose an empirical method for estimating the threshold in order to evaluate the reliability of community detection on real-world data {\em a posteriori}.
Let $\widehat{n}_i$ and $\widehat{m}_i$ denote the size and the number of edges of the identified community $i$. Define the empirical estimators
\begin{align}
\hp&=\text{number of identified external edges}/\hnone \hntwo;\\
\hp_i&=\frac{\widehat{m}_i}{\widehat{n}_i^2}; \\
\hpstar&=\sqrt{\hpone \hptwo}.
\end{align}
We apply these estimators to the political blog data in \cite{Adamic05}, where this dataset contains $1222$ blogs, labeled as either conservative or liberal, and an edge corresponds to a hyperlink reference between blogs. The detectability using the spectral modularity method is $0.9419$ (the labels are predicted by taking the sign function on the leading eigenvector of the modularity matrix). The corresponding empirical estimates are $\hp=0.0042$, $\hpone=0.0244$, $\hptwo=0.0179$, and $\hpstar=0.0209$. The high detectability of the spectral modularity method is consistent with the fact that the empirical estimate $\hp$ is below the empirical phase transition threshold $\hpstar$.

\section{Conclusion}
This paper establishes a universal phase transition threshold $p^* \asconv \sqrt{\pone \ptwo}$ on community detectability using the spectral modularity
method for a general stochastic block model.
The critical phase transition is universal in the sense that it does not depend on the community sizes. An empirical method is proposed to estimate the phase transition threshold from real-world data.

\begin{acknowledgments}
This work was supported in part by the US Army Research Office under grant W911NF-12-1-0443.
\end{acknowledgments}

\bibliographystyle{apsrev4-1}
\bibliography{IEEEabrv,PRL_modularity}
\nocite{*}

\clearpage


\appendix
\section{Proof of the fact that $\mathbb{E} \Lb \sigma_1\lb \frac{\bDelta}{\sqrt{\none \ntwo}} \rb \Rb \ra 0$ as $\none,\ntwo \ra \infty$}
\label{appen_Latala}
Since $\bDelta=\bC-\bCbar$, we have $\bDelta_{ij}=1-p$ with probability $p$ and $\bDelta_{ij}=-p$ with probability $1-p$.
Latala's theorem \cite{Latala05} states that for any random matrix $\mathbf{M}$ with statistically independent and zero mean entries, there exists a positive constant $c_1$ such that
\begin{align}
\mathbb{E} \Lb \sigma_1(\mathbf{M})\Rb \leq &c_1 \lb \max_i\sqrt{\sum_j \mathbb{E} \Lb \mathbf{M}_{ij}^2 \Rb}+\max_j\sqrt{\sum_i \mathbb{E} \Lb \mathbf{M}_{ij}^2 \Rb} \right.
\nonumber \\
 &~~~~~\left.+ \sqrt[4]{\sum_{ij} \mathbb{E} \Lb \mathbf{M}_{ij}^4 \Rb}\rb.
\end{align}
It is clear that $\mathbb{E} \Lb \bDelta_{ij} \Rb=0$ and each entry in $\bDelta$ is independent. By using $\mathbf{M}=\frac{\bDelta}{\sqrt{\none \ntwo}}$ in Latala's theorem, since $p \in [0,1]$, we have $\max_i\sqrt{\sum_j \mathbb{E} \Lb \mathbf{M}_{ij}^2 \Rb}=O(\frac{1}{\sqrt{n_1}})$, $\max_j\sqrt{\sum_i \mathbb{E} \Lb \mathbf{M}_{ij}^2 \Rb}=O(\frac{1}{\sqrt{n_2}})$, and $\sqrt[4]{\sum_{ij} \mathbb{E} \Lb \mathbf{M}_{ij}^4 \Rb}=O(\frac{1}{\sqrt[4]{\none \ntwo}})$.
Therefore $\mathbb{E} \Lb \sigma_1\lb \frac{\bDelta}{\sqrt{\none \ntwo}} \rb \Rb \ra 0$ as $\none,\ntwo \ra \infty$.

\section{Proof of (\ref{eqn_Talagrand})}
\label{appen_Talagrand}
Talagrand's concentration theorem is stated as follows. Let $g: \mathbb{R}^k \mapsto \mathbb{R}$ be a convex and 1-Lipschitz function.
Let $\bx \in \mathbb{R}^k$ be a random vector and assume that every element of $\bx$ satisfies
$|\bx_i|  \leq K$ for all $i=1,2,\ldots,k$, with probability one. Then there exist positive constants $c_2$ and $c_3$ such that for any $\epsilon >0$,
\begin{align}
\text{Pr}\lb \left| g(\bx)-\mathbb{E} \Lb g(\bx)  \Rb \right| \geq \epsilon\rb \leq c_2 \exp \lb \frac{-c_3 \epsilon^2}{K^2} \rb.
\end{align}
It is well-known that the largest singular value of a matrix $\mathbf{M}$ can be represented as $\sigma_1(\mathbf{M})=\max_{\bz^T\bz=1}||\mathbf{M} \bz||_2$ \cite{HornMatrixAnalysis} so that $\sigma_1(\mathbf{M})$ is a convex and 1-Lipschitz function.  Recall that $\bDelta_{ij}=1-p$ with probability $p$ and $\bDelta_{ij}=-p$ with probability $1-p$. Therefore applying Talagrand's theorem by substituting $\mathbf{M}=\frac{\bDelta}{\sqrt{\none \ntwo}}$ and using the facts that $\mathbb{E} \Lb \sigma_1\lb \frac{\bDelta}{\sqrt{\none \ntwo}} \rb \Rb \ra 0$ and $\frac{\bDelta_{ij}}{\sqrt{\none \ntwo}} \leq \frac{1}{\sqrt{\none \ntwo}}$, we have
\begin{align}
\text{Pr}\lb  \sigma_1 \lb \frac{\bDelta}{\sqrt{\none \ntwo}} \rb \geq \epsilon \rb \leq c_2 \exp \lb -c_3 \none \ntwo \epsilon^2 \rb.
\end{align}
Note that, since for any positive integer $\none, \ntwo >0$  $\none \ntwo \geq \frac{\none+\ntwo}{2}$,
$\sum_{\none,\ntwo} c_2 \exp \lb -c_3 \none \ntwo \epsilon^2 \rb < \infty$. Hence, by Borel-Cantelli lemma \cite{Resnick13},
$\sigma_1 \lb \frac{\bDelta}{\sqrt{\none \ntwo}} \rb \asconv 0$
when $\none, \ntwo \ra \infty$.
Finally, a standard matrix perturbation theory result \cite{HornMatrixAnalysis} is
$|\sigma_i(\bCbar+\bDelta)-\sigma_i(\bCbar)| \leq \sigma_1(\bDelta)$
for all $i$, and as $\sigma_1\lb \frac{\bDelta}{\sqrt{\none \ntwo}} \rb \asconv 0$, we have
\begin{align}
&\sigma_1\lb \frac{\bC}{\sqrt{\none \ntwo}} \rb=\sigma_1\lb \frac{\bCbar+\bDelta}{\sqrt{\none \ntwo}} \rb \asconv \sigma_1\lb \frac{\bCbar}{\sqrt{\none \ntwo}} \rb=p;~~ \nonumber \\
&\sigma_i \lb \frac{\bC}{\sqrt{\none \ntwo}} \rb \asconv 0,~~\forall i \geq 2
\end{align}
when $\none, \ntwo \ra \infty$.

\section{Proof of (\ref{eqn_modularity_eigvector_concentration_3})}
\label{appen_eigenvector_convergence}
We prove the result by showing $\frac{\yone^T \Bone \yone }{n} \asconv 0$ and $\frac{\ytwo^T \Btwo \ytwo }{n} \asconv 0$ such that
$\sqrt{\frac{n \none}{\ntwo}} \yone \asconv \pm \onenone~~\text{and}~~\sqrt{\frac{n \ntwo}{\none}} \ytwo \asconv \mp \onentwo$ due to the facts that
the vector of all ones is always in the null space of a modularity matrix and $\yone^T \onenone + \ytwo^T \onentwo=0$. We prove this statement by contradiction.
Assume $\yone$ and $\ytwo$ converge almost surely
to other vectors such that  $\frac{\yone^T \Bone \yone }{n} \ra c_4 \neq 0$  and $\frac{\ytwo^T \Btwo \ytwo }{n} \ra c_5 \neq 0$ and $c_4+c_5=0$ in order to satisfy (\ref{eqn_modularity_eigvector_concentration_2}).
By the concentration results in (\ref{eqn_A_concentration_SBM}) and (\ref{eqn_dtilde_concentration}), we have
\begin{align}
\frac{\yone^T \Bone \yone} {n} &= \frac{\yone^T \lb \Aone - b_1 \dtone \dtone^T \rb \yone} {n} \nonumber \\
&\asconv \frac{\yone^T \lb \pone \onenone \onenone^T - \frac{1}{\none^2 \pone} \cdot \none^2 \pone^2 \onenone \onenone^T   \rb \yone} {n} \nonumber \\
 &= 0,
 \end{align}
 and similarly $\frac{\ytwo^T \Btwo \ytwo} {n} \asconv 0$, which contradicts the assumption that $\frac{\yone^T \Bone \yone }{n} \asconv c_4 \neq 0$ and $\frac{\ytwo^T \Btwo \ytwo }{n} \asconv c_5 \neq 0$.
 Therefore $\sqrt{\frac{n \none}{\ntwo}} \yone \asconv \pm \onenone$ and $\sqrt{\frac{n \ntwo}{\none}} \ytwo \asconv \mp \onentwo$.

\section{The Effect of Community Size on Phase Transition}
\label{appen_finite_size}
\begin{figure*}[t!]
    \centering
    \includegraphics[width=3.7in]{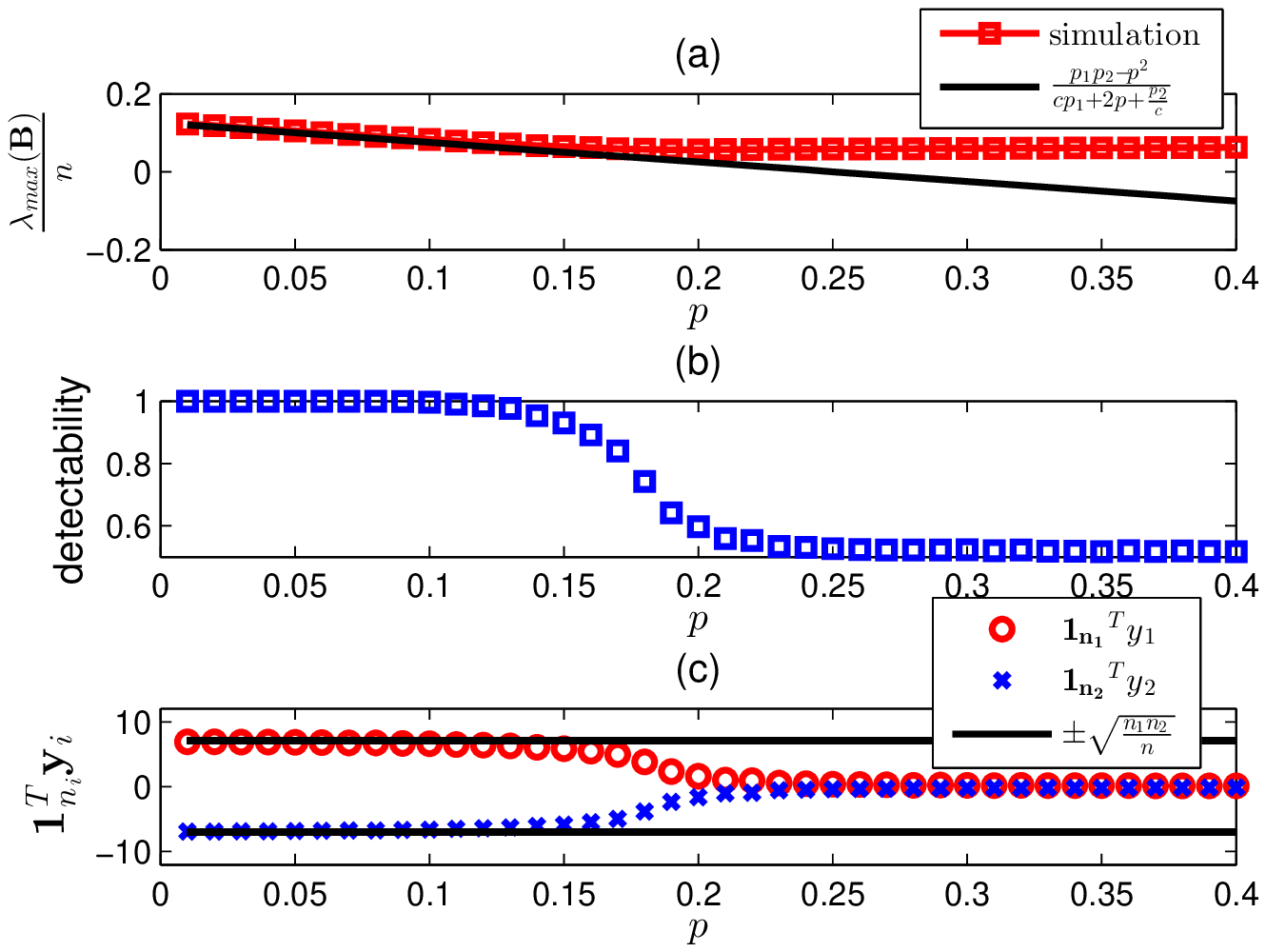}
\caption{$\none=100$, $\ntwo=100$, $\pone=0.25$, and $\ptwo=0.25$.}
\label{Fig_modularity_100_100_25_25}
\end{figure*}

\begin{figure*}[t!]
    \centering
    \includegraphics[width=3.7in]{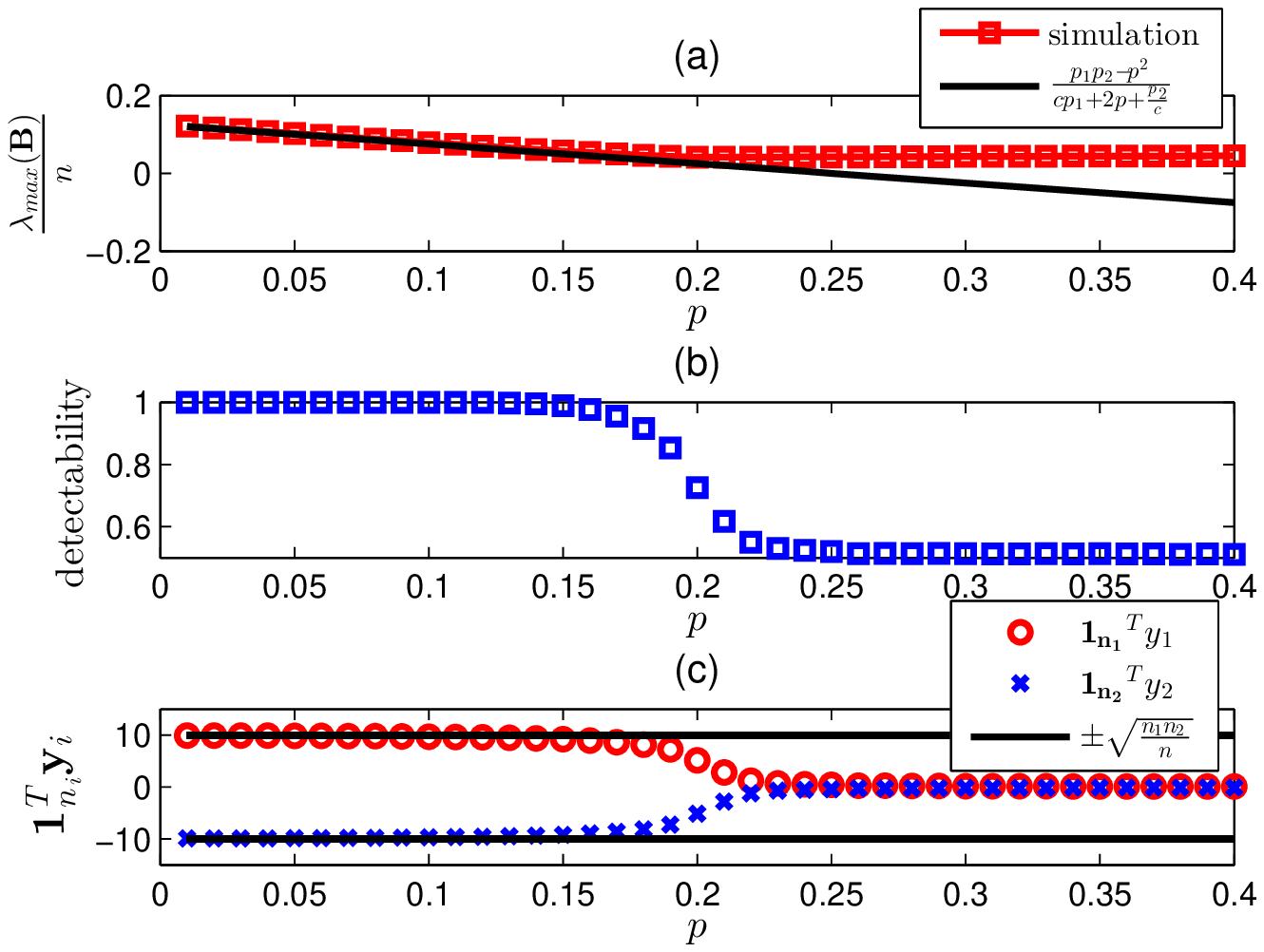}
\caption{$\none=200$, $\ntwo=200$, $\pone=0.25$, and $\ptwo=0.25$.}
\label{Fig_modularity_200_200_50_50}
\end{figure*}

\begin{figure*}[t!]
    \centering
    \includegraphics[width=3.7in]{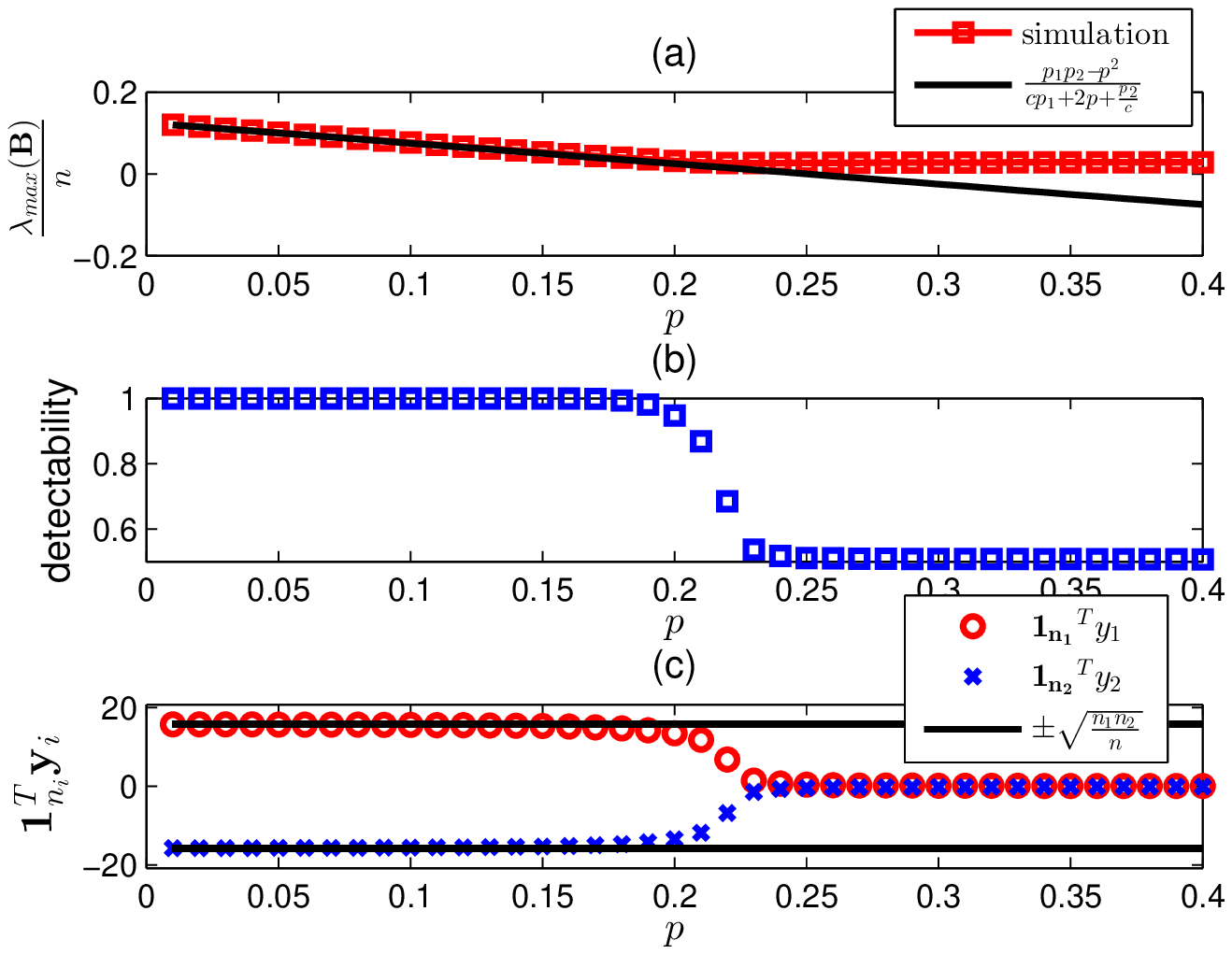}
\caption{$\none=500$, $\ntwo=500$, $\pone=0.25$, and $\ptwo=0.25$.}
\label{Fig_modularity_500_500_125_125}
\end{figure*}

\begin{figure*}[t!]
    \centering
    \includegraphics[width=3.7in]{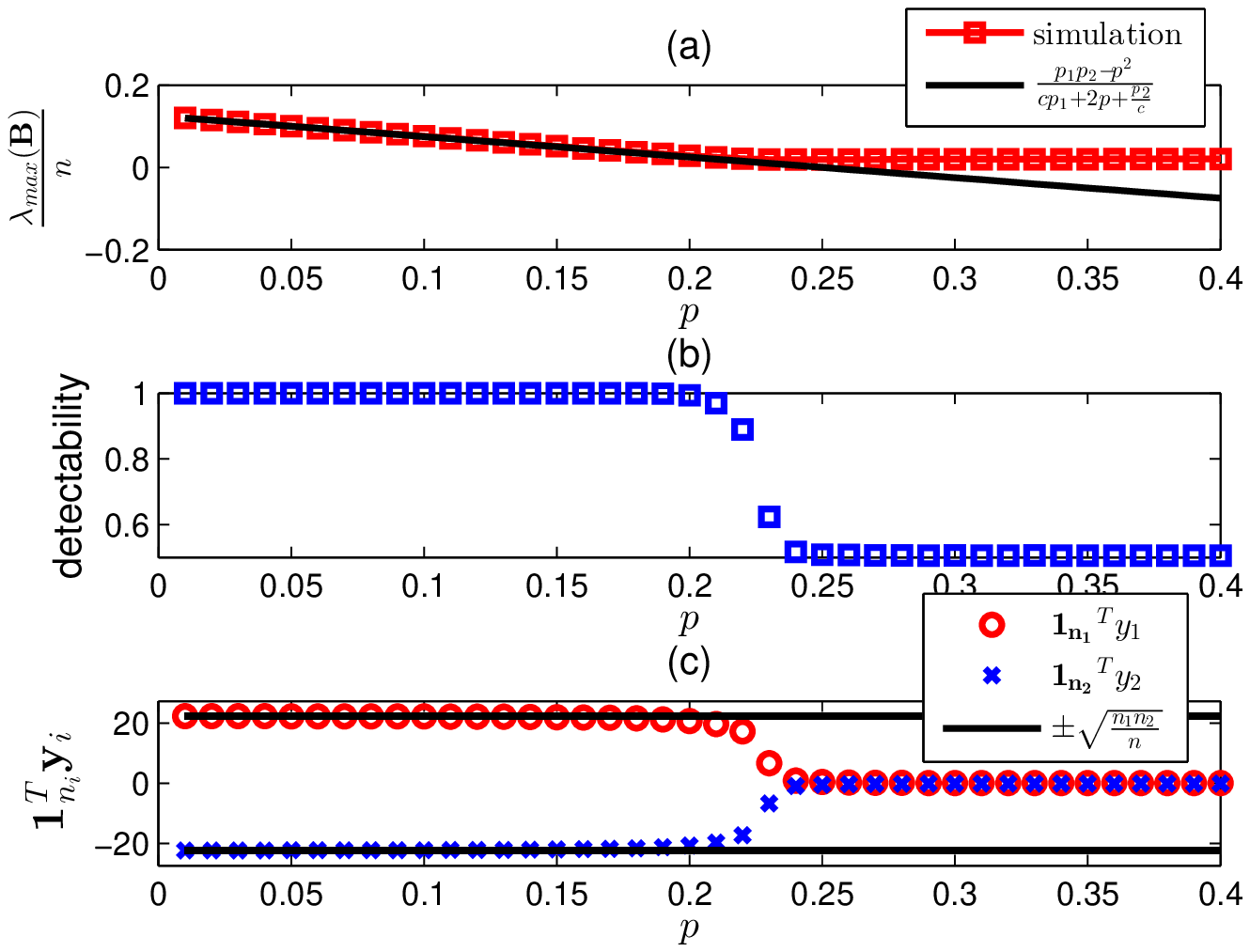}
\caption{$\none=1000$, $\ntwo=1000$, $\pone=0.25$, and $\ptwo=0.25$.}
\label{Fig_modularity_1000_1000_250_250}
\end{figure*}

\begin{figure*}[t!]
    \centering
    \includegraphics[width=3.7in]{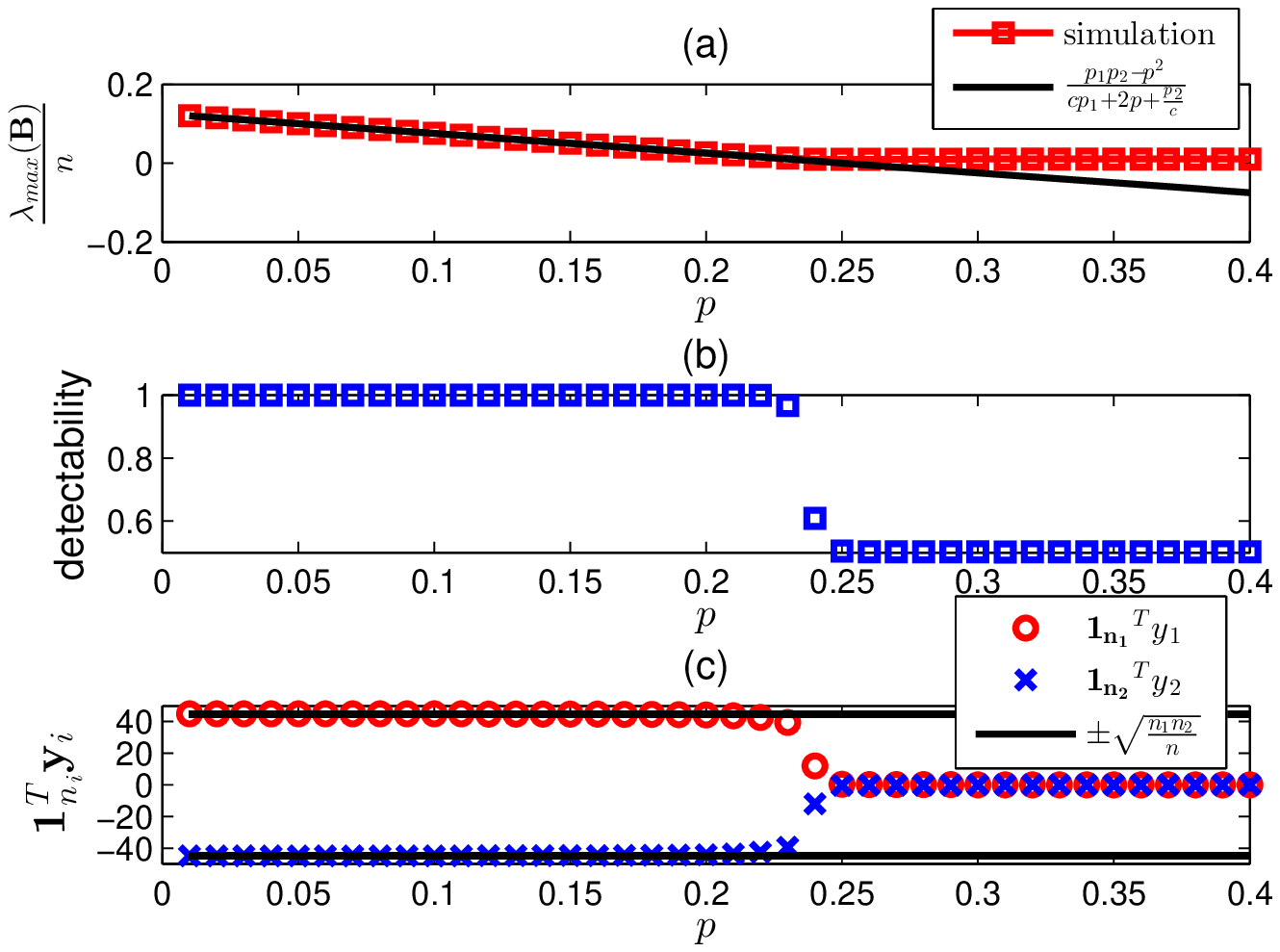}
\caption{$\none=4000$, $\ntwo=4000$, $\pone=0.25$, and $\ptwo=0.25$.}
\label{Fig_modularity_4000_4000_1000_1000}
\end{figure*}

To investigate the effect of community size on phase transition, we generate synthetic communities from the stochastic block model with different community sizes by fixing $c=1$ and $\pone=\ptwo=0.25$. The predicted phase transition threshold in (\ref{eqn_modularity_universal_phase_transition_SBM}) is $p^*=0.25$. The results (averaged for 100 runs) are shown in Fig. \ref{Fig_modularity_100_100_25_25}-\ref{Fig_modularity_4000_4000_1000_1000}. The phase transition is apparent for small community size in the sense that the spectral modularity method fails to detect the communities in the super-critical regime (i.e., the $p > p^*$ regime). In the sub-critical regime (i.e., the $p \leq p^*$ regime), we observe an intermediate regime of community detectability for small community size, and this intermediate regime vanishes as we increase the community size. This can be explained by the fluctuation of finite community size on the concentration results in (\ref{eqn_tranisition_condition_1}), (\ref{eqn_tranisition_condition_2}), (\ref{eqn_modularity_eigvector_concentration_3}), and (\ref{eqn_modularity_universal_phase_transition_SBM}). By concentration theory the fluctuation decreases with the increase of community size, and an abrupt transition occurs at the phase transition threshold $p^*$ when $\none,\ntwo \ra \infty$ and $\frac{\none}{\ntwo} \ra c >0$.


\clearpage

\end{document}